\documentclass[11pt,a4paper,english,aps,showpacs,preprint,superscriptaddress,prd]{revtex4-1}
\usepackage[latin1]{inputenc}
\usepackage{graphicx}
\usepackage{amsmath}
\usepackage{amsfonts}
\usepackage{amssymb}
\usepackage{wrapfig}
\usepackage{babel}
\usepackage{hyperref}

\begin{document}

\title{Noncommutative ${\cal N}=2$ Chern-Simons-matter model}

\author{L.~Ibiapina~Bevilaqua}
\email{leandro@ect.ufrn.br}
\affiliation{Escola de Ci\^encias e Tecnologia, Universidade Federal do Rio Grande do Norte\\
Caixa Postal 1524, 59072-970, Natal, Rio Grande do Norte, Brazil}

\author{A.~C.~Lehum}
\email{andrelehum@ect.ufrn.br}
\affiliation{Instituto de F\'{\i}sica, Universidade de S\~{a}o Paulo\\ Caixa Postal 66318, 05315-970, S\~{a}o Paulo, S\~{a}o Paulo, Brazil}
\affiliation{Escola de Ci\^encias e Tecnologia, Universidade Federal do Rio Grande do Norte\\
Caixa Postal 1524, 59072-970, Natal, Rio Grande do Norte, Brazil}


\begin{abstract}

In this work we study the three-dimensional ${\cal N}=2$ supersymmetric Chern-Simons-matter model in a noncommutative space-time. We construct the action of the noncommutative $U(N)$ non-Abelian model in terms of explicit ${\cal N}=2$ supervariables by dimensionally reducing a four-dimensional ${\cal N}=1$ supermultiplet. We also obtain the on-shell ${\cal N}=2$ supersymmetric model writing it in terms of ${\cal N}=1$ superfields. In the noncommutative Abelian case, we show that linear UV divergences are cancelled in Feynman diagrams and logarithmic divergences are absent up to one-loop order, stating that our model is free of UV/IR mixing.

\end{abstract}


\maketitle

\section{Introduction}

The noncommutativity of space-time coordinates was first suggested by Heisenberg~\cite{Jackiw:2001dj}  as a way to regulate the UV divergences present in the perturbative approach of quantum field theory. By this idea, the noncommutativity would implement an uncertainty relation to the space-time coordinates $\Delta x^m \Delta x^n=i\Theta^{mn}$, where $\Theta^{mn}$ is suggested to be of order of Planck length squared $l_p^2$ ($l_p\sim 10^{-33}$cm), introducing a minimum length, meaning that no localization is possible below such scale. Therefore, this minimum length would be taken as a natural UV cutoff accordingly to the Heisenberg's uncertainty principle ($\Delta x \Delta p\sim \hbar$). The first paper on noncommutative field theory (NCFT) was published in 1947~\cite{Snyder:1946qz}, but due to the success of the renormalization procedure this idea was forgotten until the 1990's, when it was discovered that NCFTs are obtained as low energy limit of a string theory in the presence of a Kalb-Ramond field~\cite{Seiberg:1999vs}. Another current motivation to study NCFTs is related to the ``space-time foam", i.e., at Planck length scale the space-time loses its smoothness and should involve quantum fluctuations of geometry and topology~\cite{Doplicher:1994tu}. The formulation of a NCFT would be a simpler way to implement such ideas. 

NCFTs are constructed from conventional field theories by the replacement of the ordinary product by a noncommutative one, thus all the products between fields are performed with the Moyal product~\cite{Douglas:2001ba}
\begin{equation}
\label{moyal}
f(x) * g(x) = f(x) \exp{\left[-\frac{i}{2}\overleftarrow{\partial}_m \Theta^{mn}\overrightarrow{\partial}_n \right]g(x)}
\end{equation}

\noindent where $m,n=\{0, 1, \cdots, (D-1)\}$ are space-time indices and $D$ is the dimension of the space-time. In the simplest case, $\Theta^{mn}$ is chosen to be an antisymmetric constant matrix.

Even though the noncommutativity of the space-time coordinates was first proposed to improve the UV behavior of perturbative expansion in quantum field theory, noncommutative extensions of the conventional field theory models are suitable to present the dangerous UV/IR mixing \cite{Minwalla:1999px}, that is the transmutation of part of the UV divergence of the original theory into an IR divergent behavior in its noncommutative counterpart, which could invalidate the perturbative expansion in a NCFT. A possible approach to avoid such issue is to construct noncommutative models from less UV divergent field theories. 

It is well-known that supersymmetry improve UV behavior of the field theories due to cancellations of higher order UV divergences among bosonic and fermionic contributions of the loop integrals. Therefore, supersymmetric models are natural candidates to define consistent NCFTs that are free of UV/IR mixing~\cite{Matusis:2000jf}, where the noncommutative Wess-Zumino model is a remarkable example~\cite{Girotti:2000gc}.   

In recent years, three-dimensional gauge theories have been intensely discussed in the literature because they are candidates to describe $M2$ branes~\cite{Bagger:2006sk,Krishnan:2008zm,Gustavsson:2007vu}. In particular, the three-dimensional Chern--Simons theory could be related to a topological string theory in six dimensions~\cite{Gopakumar:1998ki}. In addition, it was suggested that several ${\cal{N}}=2,3$ three-dimensional supersymmetric $U(N)$ Chern-Simons-matter models are dual to open or closed string theories in $AdS4$ in the large $N$ limit~\cite{Gaiotto:2007qi}, and its effective (super)potential has been evaluated in several approximations~\cite{effecpot}. On the other hand, non-Abelian $U(N)$ gauge field theories defined in the ordinary space-time in the large $N$ approximation are dual to noncommutative Abelian gauge theories in the large $\Theta$ limit~\cite{Ambjorn:1999ts,Aoki:1999vr,Ambjorn:2000nb,Szabo:2001kg}. This last duality suggests that the ${\cal{N}}=2,3$ noncommutative 3D Abelian Chern-Simons-matter models could also be related to string theories in $AdS4$, being possible candidates to describe $M2$ branes.

This paper is devoted to construct the noncommutative extension of the three-dimensional non-Abelian ${\cal{N}}=2$ Chern-Simons-matter model (NCCSM), being organized as follows. In Sec. \ref{n2}, following Ref.\cite{ng}, we construct the action to the noncommutative three-dimensional non-Abelian ${\cal{N}}=2$ Chern-Simons-matter model, and show how to obtain its action in terms of ${\cal N}=1$ superfields~\cite{Ivanov:1991fn}. We also integrate over Grassman coordinates and eliminate the auxiliary fields to obtain the action of the model in terms of the physical field components. In Sec. \ref{div-struc} we discuss the divergence structure of the model for the noncommutative Abelian case. In Sec. \ref{ea} we evaluate the quadratic part of the effective action to the matter and gauge superfields and show that it is free of UV/IR mixing at one-loop order. Some final remarks are done in Sec. \ref{cr}.

\section{The classical model}\label{n2}

In order to obtain the action for the ${\cal{N}}=2$ noncommutative $U(N)$ Chern-Simons theory, we start by writing the action for the $D=3$ supersymmetric Abelian Chern-Simons theory in the (usual) ${\cal N}=2$ superspace: 
\begin{equation}
\label{abelian-action}
S= \frac{1}{4}\int d^7z \; \bar{D}^\alpha V D_\alpha V = -Tr\frac{1}{4}\int d^7z \; V\; W
\end{equation}
where $z^A = (\theta^\alpha, \bar{\theta}^\alpha, x^{\alpha\beta})$ are the supercoordinates, $V=V(x,\theta,\bar{\theta})$ is the so-called prepotential (a scalar superfield that describes the gauge multiplet) and $W$  is the superfield strength
\begin{equation}
\label{W}
W = \bar{D}^\alpha D_\alpha V,
\end{equation}

\noindent according to the notations and conventions of~\cite{Gates:1983nr}, which we follow here.

The equation of motion derived from this action is obtained through the variation of $S$ with respect to $V$
\begin{equation}
\label{eqmov}
\delta S=-\frac{1}{4}\int d^7z \; (\delta V)\;  W =0 \qquad \Rightarrow \qquad W=0.
\end{equation}

Now, let us consider the noncommutative non-Abelian case. The lifting from the usual model to the noncommutative one is performed by the replacement of ll the products between fields by the Moyal one Eq.(\ref{moyal}). Since $f*g \neq g*f$ in general, we will have to the superfield strength an structure similar to the non-Abelian commutative case even in a noncommutative Abelian model. So, for the noncommutative non-Abelian theory, the superfield strength is defined by
\begin{equation}
\label{NCW}
W = \bar{D}^\alpha \left(e^{-V} D_\alpha e^V\right),
\end{equation}

\noindent where $e^V = 1+V+\frac{1}{2!} V*V +\frac{1}{3!} V*V*V + (\cdots)$, with $V=V_aT_a$ being a non-Abelian superfield ($T_a$ are the $U(N)$ group generators).

We claim that the equation of motion will have the same form as the commutative case ($W=0$) only replacing (\ref{W}) by (\ref{NCW}). In order to have this, the variation of the (still unknown) noncommutative action must be
\begin{equation}
\delta S=-Tr\frac{1}{4}\int d^7z \;(\delta V) *  W
\end{equation}

\noindent (when integrated, the Moyal product has a ciclic symmetry and when there are only two fields one may consider an ordinary commutative product, but in the action above we decided to write the star explicitly to remind us we are dealing with the noncommutative case).

On the other hand, if we write our action as $S=\int d^7z \; Tr[{\cal L}]$, its variation will be
\begin{equation}
\delta S=Tr\int d^7z \; \left[\delta V*\frac{\delta{\cal L}}{\delta V}\right].
\end{equation}

Thus, all we have to do is to solve the functional equation
\begin{equation}
\label{ncfunctionaleq}
\frac{\delta{\cal L}}{\delta V} = -\frac{1}{4} W
\end{equation}
to find the Lagrangian $\cal L$.

This can be done introducing an extra bosonic variable $t$ and defining a new superfield $\hat{V} = \hat{V}(t)$, such that 
\begin{equation}
\label{ncbc}
\hat{V}(t)\Big|_{t=0}=0 \qquad \text{and} \qquad \hat{V}(t)\Big|_{t=1}=V.
\end{equation}

For our purposes, it will be enough to take $\hat{V} = tV$, that is a simple way to satisfy the conditions above. See Nishino and Gates in \cite{ng} for an example where one does not specify the dependence on $t$, so that $\hat{V}$ is an arbitrary function of $t$, but still satisfying (\ref{ncbc}). See also \cite{vainberg} for mathematical aspects of this so-called ``Vainberg method''.

With the choice $\hat{V} = tV$, we may parametrize the variation using $t$ and take variations of the type $\delta_t \hat{V} = d\hat{V} = Vdt $, so that equation (\ref{ncfunctionaleq}) becomes an ordinary differential equation for the hatted variables
\begin{equation}
\frac{d\hat{{\cal L}}}{d\hat{V}} = \frac{d\hat{{\cal L}}}{Vdt} = -\frac{1}{4} \hat{W} \quad \Rightarrow \quad d{\hat{\cal L}} = -\frac{1}{4} V*\hat{W} dt.
\end{equation}
Because of conditions (\ref{ncbc}), we have $\hat{\cal L}(t)\Big|_{t=0}=0$ and $\hat{\cal L}(t)\Big|_{t=1}=\cal L$ and then we have
\begin{equation}
\label{nclagrangean}
{\cal L} = -\frac{1}{4}Tr\int_0^1 dt V*\hat{W} =-\frac{1}{4}Tr\int_0^1 dt V * \bar{D}^\alpha \left(e^{-\hat{V}} D_\alpha e^{\hat{V}}\right).
\end{equation}
Since $t$ is a dummy integration variable, the action does not depend on this additional parameter $t$. Its final form in terms of the prepotential $V$ is
\begin{equation}
\label{ncaction}
S = -Tr\frac{1}{4}\int d^7 z \int_0^1 dt V*\bar{D}^\alpha \left(e^{-tV} D_\alpha e^{tV}\right).
\end{equation}

It is not possible to perform the integral on $t$ exactly, since it is highly non-linear on $V$. To find the first terms, write
\begin{eqnarray}
\hat{W} & = & \bar{D}^\alpha \left(e^{-tV} D_\alpha e^{tV}\right) \nonumber \\
&=& \bar{D}^\alpha \left((1-tV+\frac{t^2}{2} V*V - \cdots) D_\alpha (1+tV+\frac{t^2}{2} V*V + \cdots)\right) \nonumber\\
&=& \bar{D}^\alpha \left((1-tV+\frac{t^2}{2} V*V - \cdots) (tD_\alpha V+\frac{t^2}{2} D_\alpha V*V+\frac{t^2}{2} V *D_\alpha V + \cdots)\right) \nonumber\\
&=& \bar{D}^\alpha \left(tD_\alpha V+\frac{t^2}{2} D_\alpha V*V +\frac{t^2}{2} V* D_\alpha V - t^2 V* D_\alpha V + \cdots\right) \nonumber\\
&=& t\bar{D}^\alpha D_\alpha V+\frac{t^2}{2} \bar{D}^\alpha(D_\alpha V*V - V *D_\alpha V) + \cdots
\end{eqnarray}
\noindent so the action (\ref{ncaction}) is
\begin{equation}
\label{firsttermsofcsaction}
S = -Tr\frac{1}{8}\int d^7 z \Big[V*\bar{D}^\alpha D_\alpha V-\frac{1}{3} V*D_\alpha V*\bar{D}^\alpha V+\frac{1}{3} D_\alpha V*V* \bar{D}^\alpha V + \cdots\Big]
\end{equation}
\noindent where we have used the cyclicity of the Moyal product (and the trace over $T^a$) under the integral sign. One interest feature common to noncommutative field theories is that even for the particular Abelian $U(1)$ gauge group the above action has the same form of the non-Abelian commutative theory found in Ref.\cite{Zupnik:1988ry} and reduces to (\ref{abelian-action}) if $\Theta^{mn}=0$. 

The matter is described by complex scalar superfields, $\Phi$ and $\bar\Phi$, and the coupling of matter with Chern-Simons superfield is performed in the same way as the non-Abelian case in four dimensions, and so the action in noncommutative space-time is
\begin{eqnarray}\label{ceq00}
S=Tr\int{d^7z}\Big{\{}
\left[-\frac{1}{4}\int_0^1 dt~V*\bar{D}^\alpha \left(e^{-tV}* D_\alpha e^{tV}\right)\right]+\bar\Phi *e^{gV}*\Phi
\Big{\}}~.
\end{eqnarray} 

Now, we will proceed as in Ref.\cite{Ivanov:1991fn} to obtain the action for the ${\cal{N}}=2$ three-dimensional noncommutative supersymmetric Chern-Simons-matter model (SCSM) written in terms of ${\cal{N}}=1$ superfields. 

Note that if we integrate out $\bar{\theta}$ in (\ref{ceq00}), the action would depend on only one fermionic variable, so it would be written in terms of ${\cal N}=1$ superfields. However, since $\theta$ and $\bar{\theta}$ are complex, in order to construct ${\cal N}=1$ superspace with real spinor variables, we define:
\begin{subequations}
\label{changeofvariables}
\begin{equation}
\theta^\alpha = \theta^\alpha_1 +i\theta^\alpha_2, \qquad \bar{\theta}^\alpha = \theta^\alpha_1 -i\theta^\alpha_2,
\end{equation}
\begin{equation}
D_\alpha = \frac{1}{2} (D_\alpha^1 -iD_\alpha^2), \qquad \bar{D}_\alpha = \frac{1}{2}  (D_\alpha^1+ i D_\alpha^2).
\end{equation}
\end{subequations}

The ${\cal N}=2$ superfield $V(x,\theta,\bar{\theta})$ can be expanded as
\begin{eqnarray}
V(x,\theta,\bar{\theta}) &=& C-i\theta^\alpha\varrho_\alpha + i\bar{\theta}^\alpha\bar{\varrho}_\alpha+\theta^2 M + \bar{\theta}^2 \bar{M}-\theta^\alpha\bar{\theta}^\beta A_{\alpha\beta}\nonumber\\
&&-i\theta^\alpha\bar{\theta}^2\zeta_\alpha + i\bar{\theta}^\alpha\theta^2\bar{\zeta}_\alpha + \bar{\theta}^2\theta^2 F~,
\end{eqnarray}

\noindent and using (\ref{changeofvariables}), we can rewrite it as
\begin{eqnarray}\label{completeV}
V(x,\theta_1,\theta_2) &=& C - i\theta_1^\alpha(\varrho_\alpha - \bar{\varrho}_\alpha) + \theta_1^2(\Upsilon+M+\bar{M})\nonumber \\ 
&& + \theta_2^\alpha\Big[(\varrho_\alpha + \bar{\varrho}_\alpha) + \theta^\beta_1\Big(iV_{\alpha\beta}+iC_{\alpha\beta}(M-\bar{M})\Big)+2\theta^2_1(\zeta_\alpha+\bar{\zeta}_\alpha)\Big] \nonumber\\
&&+ \theta_2^2\Big[(\Upsilon-M-\bar{M}) + 2i\theta^\alpha_1(\zeta_\alpha-\bar{\zeta}_\alpha)-4\theta^2_1F\Big].
\end{eqnarray}
Note that we have split the second-rank field $A_{\alpha\beta}$ into its antisymmetric and symmetric part:
\begin{eqnarray}
A_{\alpha\beta}=\frac{1}{2}(C_{\alpha\beta} \Upsilon - V_{\alpha\beta}),
\end{eqnarray}

\noindent where $C_{\alpha\beta}$ is the antisymmetric symbol used to raise and lower the spinor indices as defined in~\cite{Gates:1983nr}, $\Upsilon$ is a scalar field that can be gauged away, and the symmetric field $V_{\alpha\beta}$ is the gauge connection.

For convenience, and using our gauge freedom, we can choose $\bar{\varrho}_\alpha=\varrho_\alpha=\chi_\alpha/2$, $C=0$, $\lambda_\alpha=2(\zeta_\alpha+\bar\zeta_\alpha)$, $B=i(M-\bar{M})$ and $\Upsilon=-(M+\bar{M})$ to write (\ref{completeV}) as
\begin{equation}\label{v}
V(x,\theta_1,\theta_2) = \theta_2^\alpha\Gamma_\alpha(x, \theta_1) + \theta_2^2 H(x,\theta_1),
\end{equation}
\noindent where
\begin{subequations}
\label{N=1superfields}
\begin{equation}
\Gamma_\alpha(x,\theta_1) = \chi_\alpha + \theta^\beta_1\left(iV_{\alpha\beta}+C_{\alpha\beta}B\right)-\theta^2_1\lambda_{\alpha},
\end{equation}
\begin{equation}
H(x, \theta_1) = -2(M+\bar{M}) + \theta^\alpha_1 \left[2i(\zeta_\alpha - \bar{\zeta}_\alpha)\right] - \theta^2_1 (4F),
\end{equation}
\end{subequations}
are ${\cal N}=1$ real superfields.

We can use (\ref{changeofvariables}) and (\ref{v})  to derive
\begin{eqnarray}
V\bar{D}^\alpha D_\alpha V & = & 2 (\theta_2^\alpha\Gamma_\alpha + \theta_2^2 H )(D_1^2 + D_2^2) (\theta_2^\alpha\Gamma_\alpha + \theta_2^2 H )\nonumber\\
& = & 2 (\theta_2^\alpha\Gamma_\alpha + \theta_2^2 H )(\theta_2^\alpha D_1^2\Gamma_\alpha + \theta_2^2 D_1^2 H + H - i\theta_2^\beta\partial_\beta {}^\alpha \Gamma_\alpha)\nonumber\\
& = & 2 (\Gamma_\beta H)\theta^\beta_2 - (\Gamma_\beta D^\alpha D^\beta\Gamma_\alpha + H^2)\theta_2^2,
\end{eqnarray}
and the other terms on Eq. (\ref{firsttermsofcsaction}), to then be able to integrate over $\theta_2$ and obtain the Chern-Simons term of our model. A similar procedure could be done to $\Phi$ and $\bar{\Phi}$ to deal with the second term in (\ref{ceq00}). Moreover, we note that the auxiliary non-propagating superfield $H$ ensures off-shell ${\cal N}=2$ supersymmetry to the action, but we can use its equation of motion to get rid of it (so, we will have off-shell ${\cal N}=1$, and on-shell ${\cal N}=2$ supersymmetries).

Finally, the action to the ${\cal{N}}=2$ three-dimensional noncommutative supersymmetric Chern-Simons-matter model (NCSCSM) written in terms of ${\cal{N}}=1$ superfields can be cast as
\begin{eqnarray}\label{ceq01}
S=Tr\int{d^5z}\Big{\{}&&-\frac{1}{2}\Gamma^{\alpha}W_{\alpha}-i\frac{g}{12}\{\Gamma^{\alpha},\Gamma^{\beta}\}D_{\beta}\Gamma_\alpha-\frac{g^2}{24}\{\Gamma^{\alpha},\Gamma^{\beta}\}\{\Gamma_\alpha,\Gamma_\beta\}\nonumber\\
&&-\frac{1}{2}\overline{\nabla^{\alpha}\phi}\nabla_{\alpha}\phi +\dfrac{g^2}{4} (\bar\phi\phi)^2+\mathcal{L}_{GF}\Big{\}},
\end{eqnarray}

\noindent where  $W^{\alpha}=\frac{1}{2}D^{\beta}D^{\alpha}\Gamma_{\beta}-i\frac{g}{2}[\Gamma^\beta,D_\beta\Gamma_\alpha]-\frac{g^2}{6}[\Gamma^\beta,\{\Gamma_\beta,\Gamma_\alpha\}]$ is the gauge superfield strength with $\Gamma_{\beta}$ being the gauge superfield, $\nabla^{\alpha}=(D^{\alpha}-ig\Gamma^{\alpha})$ is the supercovariant derivative. In the above action, we have added also the corresponding Fadeev-Popov Lagrangian $\mathcal{L}_{GF}=-(D^{\alpha}\Gamma_{\alpha})^2/4\xi+\bar{c}D^{\alpha}(D_{\alpha}c-ig[\Gamma_{\alpha},c])/2$, in order to quantize the model.  

To obtain the physical content of the model, we can integrate over the remaining Grassmann variables. Let us consider the superfield expansions as in the Appendix \ref{components}. In the Wess-Zumino gauge $\chi=B=0$, the action Eq.(\ref{ceq01}) can be cast as  
\begin{eqnarray}\label{ceq01a}
S&=&Tr\int{d^3x}\Big{\{}-\frac{1}{4}\lambda^{\alpha}\lambda_{\alpha}-\frac{1}{2}V^{\alpha\gamma}i{\partial_\gamma}^{\beta}V_{\beta\alpha}+\frac{g}{6}{V^\alpha}_{\beta}[V^{\beta\gamma},V_{\gamma\alpha}]+\bar{F}F+\bar\varphi\Box\varphi+\bar\psi^\beta i{\partial_\beta}^{\alpha}\psi_\alpha\nonumber\\
&&-\frac{g^2}{2}\bar\varphi V^{\alpha\beta}V_{\alpha\beta}\varphi-i\frac{g}{2}\left(\bar\varphi V^{\alpha\beta}\partial_{\alpha\beta}\varphi
-\partial_{\alpha\beta}\bar\varphi V^{\alpha\beta}\varphi+2\bar\psi^\alpha i{V_\alpha}^{\beta}\psi_\beta+\bar\varphi\lambda^{\alpha}\psi_\alpha-\bar\psi^{\alpha}\lambda_\alpha\varphi \right)\nonumber\\
&&+\frac{g^2}{2}\left(\bar{F}\varphi\bar\varphi\varphi+\bar\varphi F\bar\varphi\varphi+\bar\psi^{\alpha}\psi_{\alpha}\bar\varphi\varphi
+\psi^{\alpha}\bar\psi_{\alpha}\varphi\bar\varphi+\frac{1}{2}\bar\psi^\alpha\varphi\bar\psi_\alpha\varphi+\frac{1}{2}\bar\varphi\psi^\alpha\bar\varphi\psi_\alpha \right)\Big{\}}.
\end{eqnarray}

The superpartner $\lambda$ of the Chern-Simons field $V^{\alpha\beta}$ is not a dynamical field and can be integrated out (together with the auxiliary fields $F$ and $\bar{F}$), resulting
\begin{eqnarray}\label{ceq01b}
S&=&Tr\int{d^3x}\Big{\{}-\frac{1}{2}V^{\alpha\gamma}i{\partial_\gamma}^{\beta}V_{\beta\alpha}+\frac{g}{6}{V^\alpha}_{\beta}[V^{\beta\gamma},V_{\gamma\alpha}]+\bar\varphi\Box\varphi+\bar\psi^\beta i{\partial_\beta}^{\alpha}\psi_\alpha\nonumber\\
&&-i\frac{g}{2}\left(\bar\varphi V^{\alpha\beta}\partial_{\alpha\beta}\varphi-\partial_{\alpha\beta}\bar\varphi V^{\alpha\beta}\varphi+2\bar\psi^\alpha i{V_\alpha}^{\beta}\psi_\beta+\bar\varphi\lambda^{\alpha}\psi_\alpha-\bar\psi^{\alpha}\lambda_\alpha\varphi \right)\nonumber\\
&&-\frac{g^2}{2}\bar\varphi V^{\alpha\beta}V_{\alpha\beta}\varphi 
+\frac{g^2}{2}\left(\bar\psi^{\alpha} \varphi \bar\varphi\psi_{\alpha}+2\bar\psi^{\alpha}\psi_{\alpha}\bar\varphi\varphi-\frac{1}{2}(\bar\varphi\varphi)^3\right)\Big{\}}~.
\end{eqnarray}

\section{Divergences structure}\label{div-struc}

Hereafter, in order to study the structure of UV divergences and UV/IR mixing of the model, let us choose the particular Abelian $U(1)$ noncommutative gauge group. The noncommutative vertices are evaluated in the Appendix \ref{vertices} and from the quadratic part of the Eq.(\ref{ceq01}) we obtain the following propagators
\begin{eqnarray}
&&\langle\Gamma_{\alpha}(p,\theta_1)\Gamma_{\beta}(-p,\theta_2)\rangle=-\frac{i}{2p^2}(D_{\beta}D_{\alpha}-\xi D_{\alpha}D_{\beta})\delta^{(2)}(\theta_1-\theta_2);\nonumber\\
&&\langle\phi(p,\theta_1)\bar\phi(-p,\theta_2)\rangle=\frac{i}{p^2}D^2\delta^{(2)}(\theta_1-\theta_2);\\
&&\langle c(p,\theta_1)\bar{c}(-p,\theta_2)\rangle=\frac{i}{p^2}D^2\delta^{(2)}(\theta_1-\theta_2).\nonumber
\end{eqnarray}

Let us compute the superficial degree of divergence $\omega$ of a given diagram $\mathcal{F}$. It is a guide to understand where UV divergences and, therefore, UV/IR mixings are suitable to appear. First, each loop contributes to $\omega$ with $2$ ($3$ for the momentum integration and $-1$ from the contraction of the loop at one point, that according to superspace properties it is necessary an operator $D^2$ be applied over a $\delta(\theta)$, which could be converted to momentum with a power $1$). We know that each super-derivative operator $D$ contribute with a power one-half of momentum to a given diagram, therefore these operators contribute with a factor of $1/2$ to $\omega$, and we see that each propagator of the model behaves like $1\over p$ in momentum scale, contributing with a factor of $-1$. The vertices $V_c$, $V_d$ and $V_f$, which possess a derivative $D$, contribute with a factor of $1/2$. So, the superficial degree of divergence $\omega$ of the model can be cast as
\begin{eqnarray}\label{omega1}
\omega=2L-P_\phi-P_\Gamma-P_c+\frac{V_c+V_d+V_f}{2}-\frac{N_D}{2},
\end{eqnarray}

\noindent where $L$ stands for the number of loops of $\mathcal{F}$, $N_D$ in the number of $D$ operators applied in the external legs, and $P_\phi$, $P_\Gamma$, and $P_c$ are the number of matter, gauge and ghost superfield propagators, respectively. 

Employing the topological relation $L+V-P=1$, $\omega$ becomes
\begin{eqnarray}\label{omega2}
\omega=2+P_\phi+P_\Gamma+P_c-2V_a-2V_b-2V_e-\frac{3}{2}(V_c+V_d+V_f)-\frac{N_D}{2}.
\end{eqnarray}

The number of the propagators $P_j$ in $\mathcal{F}$ is related to the number of the superfields $N_j$ (of type $j$) used to construct it and the number of external legs $E_j$ by $P_j=\frac{1}{2}(N_j-E_j)$. It is easy to see that the $N_j$ in $\mathcal{F}$ is related to the vertices by
\begin{eqnarray}\label{omega4}
N_\phi&=&4V_a+2V_b+2V_c;\nonumber\\
N_\Gamma&=&2V_b+V_c+3V_d+4V_e+V_f;\nonumber\\
N_c &=&2V_f.\nonumber
\end{eqnarray}

\noindent Substituting these relations into Eq.(\ref{omega2}), we finally obtain
\begin{eqnarray}\label{omega4}
\omega=2-\frac{E_\phi}{2}-\frac{E_\Gamma}{2}-\frac{N_D}{2}.
\end{eqnarray}

The first lesson we have from the superficial degree of divergence is that quadratic divergences are suitable to appear only in vacuum diagrams, i.e., diagrams without any external legs (such diagrams are important to evaluate the effective action through the background-field formalism), but they are not relevant in scattering processes, for example. It is easy to see that any diagram with more than four external superfields are UV finite. 

Linear divergences can appear in the self-energy diagrams of $\phi$ and $\Gamma$. Actually, we will show in the next section that these diagrams are UV finite and free of UV/IR mixing at one-loop order. Moreover, The self-energy diagrams to the $\Gamma$ superfield is protected by the gauge symmetry, therefore they have to present at least $2$ derivatives applied to the external legs, like a Chern-Simons term $\Gamma DD\Gamma$, still allowing a logarithmic UV divergence. But the extended supersymmetry ${\cal{N}}=2$ protects this term from loop corrections, so the minor number of $D$'s applied in the external legs allowed in self-energy diagrams to the gauge superfied is $4$ (generating a Maxwell-like term $DD\Gamma DD\Gamma$), turning such process UV finite and free of UV/IR mixing to all loops. A similar argument can be used to fix the UV behavior of the matter superfield self-energy diagrams. The extended supersymmetry ${\cal{N}}=2$ and the superconformal invariance protect the such processes from mass renormalization, only allowing a wave function one. Therefore to $\phi$ self-energy diagrams we have to consider at least two $D$'s applied to the external legs, consequently only a logarithmic divergence is allowed in such processes.  

Logarithmic divergences at one-loop order in three-dimensional space-time should appear as an odd power of the integration momentum in the numerator, e.g. $\int{d^3k} (p\cdot k)/(k^2+m^2)^2$, which is obviously vanishing. Therefore, at one-loop level, any diagram with more than two external legs are UV finite. Another possibility of logarithmic UV divergences comes from the three and four-point functions of the gauge superfield, as $\Gamma\Gamma D\Gamma$ and $\Gamma\Gamma\Gamma\Gamma$, and the four-point functions of the matter superfield $\bar\phi\phi\bar\phi\phi$. Since the logarithmic divergences are absent at one-loop order, such divergences can only occur in higher orders in the perturbative expansion. Anyway, these type of divergences are not suitable to present UV/IR mixings~\cite{Matusis:2000jf}. 

Next section we will show how the linear UV divergences are cancelled in Feynman graphs at one-loop order involving two external legs of matter and gauge superfields, result which suggests that the noncommutative ${\cal{N}}=2$ Chern-Simons-matter model is free of UV/IR mixing. 

\section{Effective action}\label{ea}

In the last section we discussed how UV divergences and UV/IR mixings can appear in the NCSCSM. The only source of UV/IR mixing are the two-point functions of the matter and gauge superfields. Therefore, let us compute the quadratic part of the effective action for $\phi$ and $\Gamma$. To do this, we will use the action of the model written in term of ${\cal{N}}=1$ superfields, Eq.(\ref{ceq01}). The noncommutative vertices are given in the Appendix \ref{vertices}. 

\begin{figure}[ht]
 \begin{center}
\includegraphics[width=10cm]{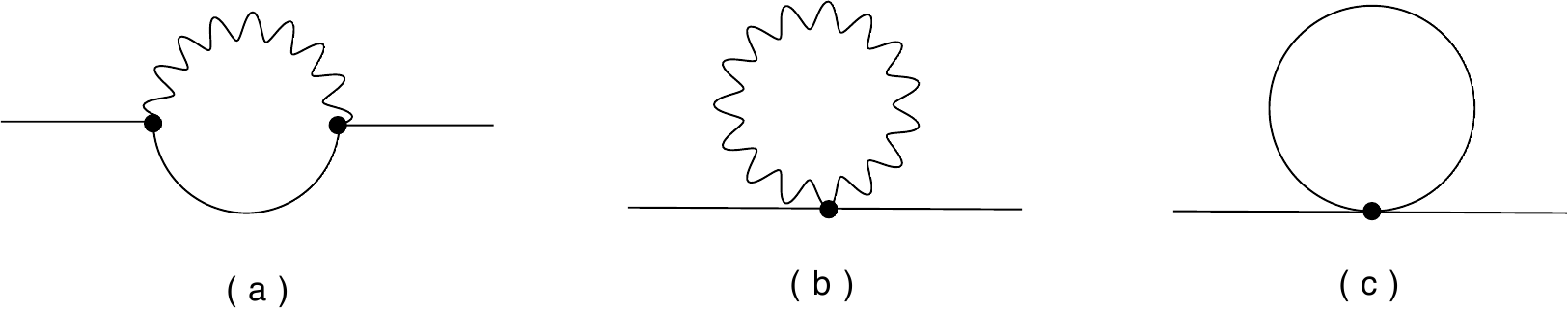}
  \end{center}
\caption{One-loop contributions to the quadratic part of the matter superfield effective action. Wavy and continuos lines represent the gauge and matter superfield propagators, respectively.} \label{fig1}
\end{figure}

The Feynman diagrams to the one-loop contribution to the effective action for the matter superfields are drawn in Figure~\ref{fig1}. The corresponding expression to the diagram \ref{fig1}(a) can be cast as
\begin{eqnarray}\label{ceq1}
\Gamma^{(2)}_{\ref{fig1}a}=\int{\frac{d^3p}{(2\pi)^3}}d^2\theta~\bar\phi(p,\theta)~\left[\int\frac{d^3k}{(2\pi)^3}\frac{-3g^2(1+\xi)p^2+\xi k^2}{k^2(k-p)^2} \right]\phi(-p,\theta),
\end{eqnarray}

\noindent while the \ref{fig1}(b) and \ref{fig1}(c) contributions are given by
\begin{eqnarray}\label{ceq2}
\Gamma^{(2)}_{\ref{fig1}b}=\int{\frac{d^3p}{(2\pi)^3}}d^2\theta~\bar\phi(p,\theta)\left[\int\frac{d^3k}{(2\pi)^3}\frac{g^2(1+\xi)}{k^2} \right]\phi(-p,\theta),
\end{eqnarray}

\noindent and
\begin{eqnarray}\label{ceq3}
\Gamma^{(2)}_{\ref{fig1}c}=-\int{\frac{d^3p}{(2\pi)^3}}d^2\theta~\bar\phi(p,\theta)\left[\int\frac{d^3k}{(2\pi)^3}\frac{g^2}{k^2} \right]\phi(-p,\theta),
\end{eqnarray}

\noindent respectively. 

The last term of Eq.(\ref{ceq1}), and the expressions Eq.(\ref{ceq2}) and Eq.(\ref{ceq3}) are linearly UV divergent. Summing up the three diagrams, we have
\begin{eqnarray}\label{ceq4}
\Gamma^{(2)}_{matter}=-g^2(1+\xi)\int{\frac{d^3p}{(2\pi)^3}}d^2\theta~\bar\phi(p,\theta)~p^2\phi(-p,\theta)\left[\int\frac{d^3k}{(2\pi)^3}\frac{1}{k^2(k-p)^2} \right].
\end{eqnarray}
 
\noindent The final result is UV finite and independent of the noncommutative parameter $\Theta$. In particular, we can see that the one-loop correction to the scalar superfield propagator is a higher derivative term, that at lowest order in powers of $p^2$ should lead to the Lagrangian $\mathcal{L}_{\bar\phi\phi}\propto\bar\phi\Box\phi$.

\begin{figure}[ht]
 \begin{center}
\includegraphics[width=12cm]{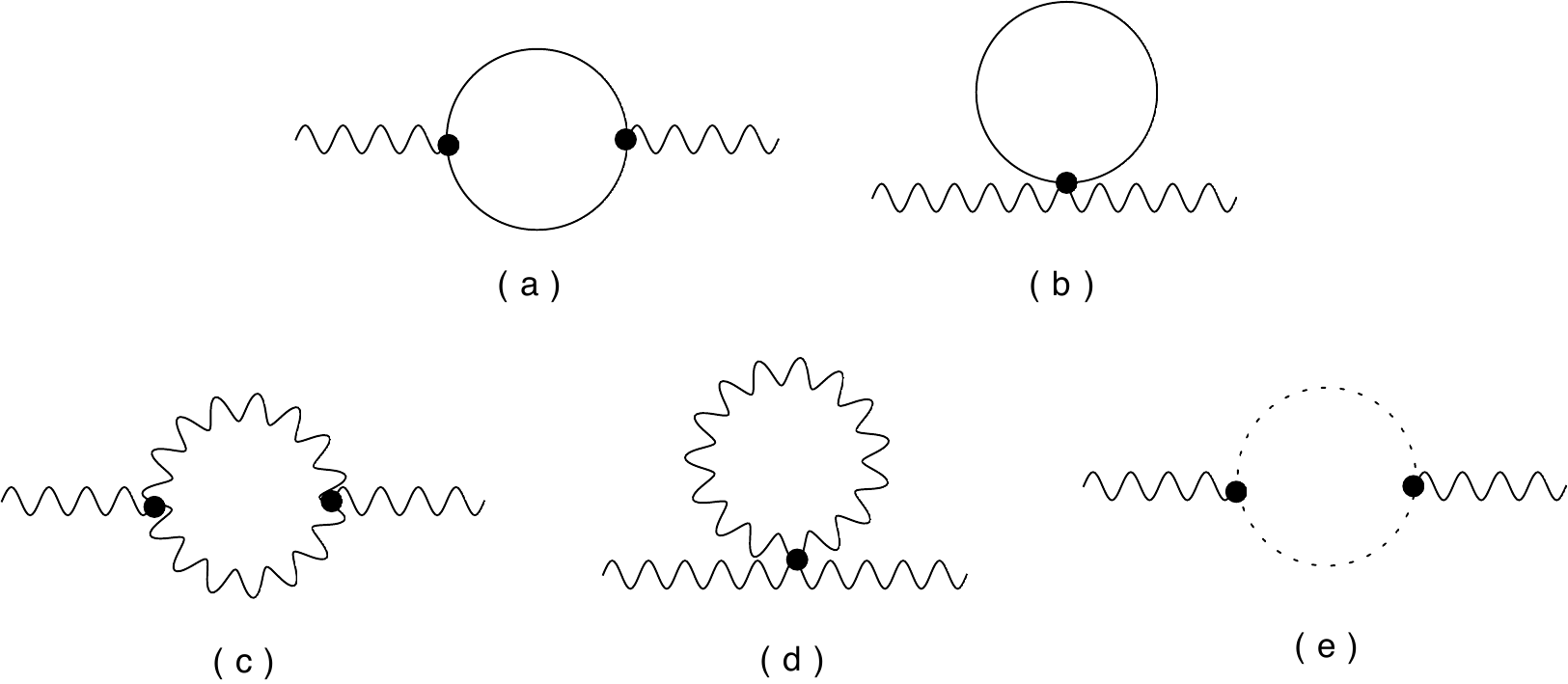}
  \end{center}
\caption{ One-loop contributions to the quadratic part of the gauge superfield effective action. Wavy, continuous and dashed lines represent the gauge, matter and ghost superfield propagators, respectively.}  \label{fig2}
\end{figure}

The quadratic part of the gauge superfield effective action receives contributions from matter (Figures \ref{fig2} (a) and (b)) and pure gauge (Figures \ref{fig2} (c), (d) and (e)) sectors. The one-loop contributions which come from matter sector can be cast as
\begin{eqnarray}
\Gamma^{(2)}_{\ref{fig2}(a)}&=&\frac{g^2}{8}\int{\frac{d^3p}{(2\pi)^3}}d^2\theta~\Gamma^{\alpha}(p,\theta)\left[\int\frac{d^3k}{(2\pi)^3}\frac{p_{\alpha\beta} D^2+3C_{\beta\alpha}p^2+4C_{\beta\alpha}k^2}{k^2(k-p)^2} \right]\Gamma^{\beta}(-p,\theta);\label{ceq4}\\
\Gamma^{(2)}_{\ref{fig2}(b)}&=&-\frac{g^2}{2}\int{\frac{d^3p}{(2\pi)^3}}d^2\theta~\Gamma^{\alpha}(p,\theta)\left[\int\frac{d^3k}{(2\pi)^3}\frac{1}{k^2} \right]\Gamma_{\alpha}(-p,\theta).\label{ceq5}
\end{eqnarray}

Adding these two contributions, we have
\begin{eqnarray}
\Gamma^{(2)}_{\ref{fig2}(a)+(b)}&=&\frac{g^2}{8}\int{\frac{d^3p}{(2\pi)^3}}d^2\theta~\Gamma^{\alpha}(p,\theta)\left[\int\frac{d^3k}{(2\pi)^3}\frac{3C_{\beta\alpha}p^2+p_{\beta\alpha}D^2}{k^2(k-p)^2} \right]\Gamma^{\beta}(-p,\theta),\label{ceq6}
\end{eqnarray}

\noindent that is UV finite. 

The contributions coming from pure gauge sector, Figures \ref{fig2} (c), (d) and (e), are given by
\begin{eqnarray}
\Gamma^{(2)}_{\ref{fig2}(c)}&=&\frac{g^2}{24}\int{\frac{d^3p}{(2\pi)^3}}d^2\theta~\Gamma^{\alpha}(p,\theta)\nonumber\\
&&\left[\int\frac{d^3k}{(2\pi)^3}\frac{(1+\xi)^2p_{\alpha\beta} D^2+C_{\beta\alpha}(-\xi^2p^2+\frac{5}{6}p^2+12\xi k^2)}{k^2(k-p)^2} \right]\Gamma^{\beta}(-p,\theta);\label{ceq7}\\
\Gamma^{(2)}_{\ref{fig2}(d)}&=&-\frac{g^2}{2}(1+\xi)\int{\frac{d^3p}{(2\pi)^3}}d^2\theta~\Gamma^{\alpha}(p,\theta)\left[\int\frac{d^3k}{(2\pi)^3}\frac{C_{\beta\alpha}\sin^2(k\wedge p)}{k^2} \right]\Gamma^{\beta}(-p,\theta);\label{ceq8}\\
\Gamma^{(2)}_{\ref{fig2}(e)}&=&-\frac{g^2}{4}\int{\frac{d^3p}{(2\pi)^3}}d^2\theta~\Gamma^{\alpha}(p,\theta)\left[\int\frac{d^3k}{(2\pi)^3}\frac{(p^2-2k^2)C_{\beta\alpha}\sin^2(k\wedge p)}{k^2(k-p)^2} \right]\Gamma^{\beta}(-p,\theta),\label{eq09}
\end{eqnarray}

\noindent which are UV divergent one-by-one, being a source of UV/IR mixing. 

However, adding these three diagrams we obtain a complete cancellation of the potentially dangerous terms, yielding the UV finite and free of UV/IR mixing effective action 
\begin{eqnarray}
\Gamma^{(2)}_{\ref{fig2}(c)+(d)+(e)}&=&\frac{g^2}{24}\int{\frac{d^3p}{(2\pi)^3}}d^2\theta \int\frac{d^3k}{(2\pi)^3}\frac{\sin^2(k\wedge p)}{k^2(k-p)^2}\nonumber\\
&&\Gamma^{\alpha}(p,\theta)\left[ (1+\xi)^2p_{\alpha\beta} D^2+(1-\xi^2)C^{\alpha\beta}p^2\right] \Gamma^{\beta}(-p).\label{ceq10}
\end{eqnarray}

We see that the quadratic part of the one-loop effective action, Eqs. (\ref{ceq6}) and (\ref{ceq10}), is the generation of a Maxwell-like term. There is no correction to the Chern-Simons one, $\Gamma^{\alpha}D^{\beta}D_{\alpha}\Gamma_{\beta}$, because the three-dimensional $\mathcal{N}=2$ extension can be viewed as a result of a dimensional reduction from the four-dimensional $\mathcal{N}=1$ model, where the fermions are chiral. Therefore, the original symmetries of the model prevent the one-loop correction to the Chern-Simons term.  

Notice that there is no wave function renormalization at one loop order, even a finite one, since the Chern-Simons receives no one-loop contribution and the one-loop effective action to the matter superfield, Eq.(\ref{ceq6}), only generates a higher order derivative term
\begin{eqnarray}
\int{d^2\theta}\bar\phi\Box\phi = c_1 \bar{\varphi}\Box^2 \varphi+c_2\bar\psi^{\alpha}{\partial_{\alpha}}^{\beta}\Box\psi_{\beta},
\end{eqnarray}

\noindent where in the right side of the last equation is obtained after elimination of the auxiliary field $F$, $c_1$ and $c_2$ are constants. 

Even though we stated that N2NCCSM is free of UV/IR mixing only for the Abelian case, we have reasons to believe that the non-Abelian case is also free from that dangerous behaviors, because apart from the non-Abelian group structure constants, the noncommutative Abelian Feynman rules are very similar to its non-Abelian extension.

\section{Concluding remarks}\label{cr}

In this work, we obtained the action of the noncommutative $U(N)$ ${\cal{N}}=2$ supersymmetric Chern-Simons-matter model in a three-dimensional space-time, in terms of explicit ${\cal{N}}=2$ supervariables, from four-dimensional supermultiplets through dimensional reduction method. We also obtained the action of the model in terms of ${\cal{N}}=1$ superfields and (to the particular $U(1)$ case) used it to evaluate the quadratic part of the effective action at one-loop, showing (by general arguments) that the model is free of UV/IR mixing.

This work is a step towards the finiteness of the model. We showed that the Chern-Simons term does not receive any one-loop correction, even a finite one. Moreover, we can conclude, due to a non-renormalization theorem of the Chern-Simons coupling~\cite{Sakamoto:1999cx,Brandt:2000yk,Das:2001kf}, that it does not receive any higher order corrections, and therefore the gauge (Chern-Simons) sector of the model is UV finite and free of UV/IR mixing to all loop orders.

An interesting development of our analysis would be to study the Ward identities in order to see if the finiteness of the Chern-Simons sector implies the finiteness of the full model.

Another further direction has to do with the way we introduced the noncommutativity to our supersymmetric model. Recall that we have used the superfield formalism as a convenient way to deal with supersymmetry, so we have constructed our model upon superspace, which means we have added Grassmannian coordinates to the usual space-time, ending up with the supercoordinates $z^A=\{x^m, \theta^\alpha\}$. However, the noncommutativity was introduced only for the bosonic coordinates  $x^m$, since the Moyal product (\ref{moyal}) induces the relation $[x^m,x^n] = i\Theta^{mn}$, but leaves untouched the anticommutation relation for the fermionic variables: $\{\theta^\alpha, \theta^\beta\} =0$. 

A more general approach would be to consider a suitable generalized Moyal product for the superfields involving the full supercoordinates such that a nonanticommutativity is implied to the fermionic variable:
\begin{eqnarray}
\{\theta^\alpha, \theta^\beta\} = i \Sigma^{\alpha\beta}.
\end{eqnarray}

As noted in the introduction to our work, noncommutativity appears naturally as a low limit of string theory. We should recall that this nonanticommutativity is also indicated by string theory as shown in a later paper by Seiberg~\cite{Seiberg:2003}. A lot of work has been done to study supersymmetric non(anti)commutative models~\cite{Ferrari:2006,Klemm:2003,Grisaru:2004,Grisaru:2006}. 

It has been pointed out in~\cite{Ferrari:2006} that non(anti)commutativity can be treated in a way to avoid the UV/IR mixing in three-dimensional supersymmetric field theories. We should keep in mind, however, that the non(anti)commutativity violates the super-Poincar\`{e} algebra, although there is an elegant way to handle the symmetries of the model as a deformed supersymmetric algebra, the so called twisted-superPoincar\`{e}~\cite{Kobayashi:2005,Chaichian:2004,Ihl:2006}.

\vspace{1cm}

{\bf Acknowledgments.} This work was partially supported by the Brazilian agencies Funda\c{c}\~{a}o de Amparo \`{a} Pesquisa do Estado de S\~{a}o Paulo (FAPESP), Conselho Nacional de Desenvolvimento Cient\'{\i}fico e Tecnol\'{o}gico (CNPq) and Funda\c{c}\~{a}o de Apoio  \`{a} Pesquisa do Rio Grande do Norte (FAPERN). The authors would like to thank A. F. Ferrari for useful comments. L.I.B. also thanks Department of Mathematical Physics of Institute of Physics of University of S\~{a}o Paulo for the hospitality.

\appendix

\section{Superfield expansions}\label{components}

The superfields $\phi$, $\bar\phi$ and $\Gamma$ can be expressed in a terminating Taylor series of the Grassmanian coordinate $\theta$. Their expansions are given by
\begin{eqnarray}\label{b01}
\phi&=&\varphi+\theta^\alpha\psi_\alpha-\theta^2F,\nonumber\\
\bar\phi&=&\bar\varphi+\theta^\alpha\bar\psi_\alpha-\theta^2\bar{F},\nonumber\\
\Gamma_{\alpha}&=&\chi_{\alpha}+\theta^{\beta}\left(C_{\beta\alpha}B+iV_{\beta\alpha}\right)-\theta^{2}\lambda_{\alpha}.\nonumber
\end{eqnarray}

Due to gauge arbitrariness, we can choose the components $\chi$ and $B$ of the gauge superfield to be vanishing. This choice of gauge is known as Wess-Zumino gauge. 

\section{Noncommutative vertices}\label{vertices}

In the canonical noncommutativity ($\Theta^{mn}$ is a constant antisymmetric matrix), all information that a particle is propagating in a noncommutative space-time is due to the vertices of interaction, since the propagators are like the commutative ones. The noncommutative vertices are characterized by the presence of phases depending on noncommutative parameter $\Theta$ and the momenta flowing to the vertex. In the present model, the vertices are given by     
\begin{eqnarray}\label{a01}
V_a=\frac{g^2}{4}~\mathrm{e}^{-i[k_2\wedge(k_3+k_4)+k_3\wedge k_4]}\bar\phi(k_1)\phi(k_2)\bar\phi(k_3)\phi(k_4)~,
\end{eqnarray}
\begin{eqnarray}\label{a02}
V_b=-\frac{g^2}{2}~\mathrm{e}^{-i[k_2\wedge(k_3+k_4)+k_3\wedge k_4]}\bar\phi(k_1)\phi_2(k_2)\Gamma^{\alpha}(k_3)\Gamma_\alpha(k_4)~,
\end{eqnarray}
\begin{eqnarray}\label{a03}
V_c=\frac{ig}{2}~\mathrm{e}^{-i k_2\wedge k_3} \left[D^{\alpha}\phi(k_1)\bar\phi(k_2)\Gamma_{\alpha}(k_3)-\phi(k_1)D^{\alpha}\bar\phi(k_2)\Gamma_{\alpha}(k_3)\right]~,
\end{eqnarray}
\begin{eqnarray}\label{a04}
V_d=\frac{g}{3}\sin(k_2\wedge k_3) \Gamma^{\alpha}(k_1)\Gamma^{\beta}(k_2)D_{\beta}\Gamma_{\alpha}(k_3)~,
\end{eqnarray}
\begin{eqnarray}\label{a05}
V_e=\frac{g^2}{6}~\sin{(k_4\wedge k_3)}\sin{[k_2\wedge(k_3+k_4)]}~
\Gamma^{\alpha}(k_1)\Gamma^{\beta}(k_2)\Gamma_{\beta}(k_3)\Gamma_{\alpha}(k_4)~,
\end{eqnarray}
\begin{eqnarray}\label{a05}
V_f=g~\sin{(k_3\wedge k_2)}\bar{c}(k_1)D^{\alpha}[\Gamma_{\alpha}(k_2)c(k_3)]~,
\end{eqnarray}

\noindent where $k\wedge p=\Theta^{mn}k_m p_n/2$.

\end{document}